\documentclass[runningheads]{llncs}
\usepackage[T1]{fontenc}
\usepackage{graphicx}
\usepackage{enumitem} % 用于自定义列表样式
\usepackage{amsmath, amssymb}
\usepackage{array} % 提供增强的表格功能
\usepackage{booktabs} % 提供美观的表格线条
\begin{document}
\title{UltraDfeGAN: Detail-Enhancing Generative Adversarial Networks for High-Fidelity Functional Ultrasound Synthesis}
\titlerunning{Detail-Enhancing Generative Adversarial Networks}
\author{Zhuo Li\inst{1,2} \and
Xuhang Chen\inst{1} \and
Shuqiang Wang\inst{1,2}\orcidID{0000-0003-1119-320X} \and
Bin Yuan\inst{1} \and
Nou Sotheany \inst{3} \and
Ngeth Rithea \inst{4}}

\authorrunning{Zhuo Li et al.}

\institute{Shenzhen Institute of Advanced Technology, Chinese Academy of Sciences, Shenzhen 518055, China \and University of Chinese Academy of Sciences, beijing,100049, china \and Department of Information and Communication, Institute of Technology of Cambodia, Phnom Penh, Cambodia \and Department of Telecommunication and Network Engineering, Institute of Technology of Cambodia, Phnom Penh, Cambodia\\
\email{sq.wang@siat.ac.cn}
}

\maketitle              % typeset the header of the contribution
\begin{abstract}
Functional ultrasound (fUS) is a neuroimaging technique known for its high spatiotemporal resolution, enabling non-invasive observation of brain activity through neurovascular coupling. Despite its potential in clinical applications such as neonatal monitoring and intraoperative guidance, the development of fUS faces challenges related to data scarcity and limitations in generating realistic fUS images. This paper explores the use of a generative adversarial network (GAN) framework tailored for fUS image synthesis. The proposed method incorporates architectural enhancements, including feature enhancement modules and normalization techniques, aiming to improve the fidelity and physiological plausibility of generated images. The study evaluates the performance of the framework against existing generative models, demonstrating its capability to produce high-quality fUS images under various experimental conditions. Additionally, the synthesized images are assessed for their utility in downstream tasks, showing improvements in classification accuracy when used for data augmentation. Experimental results are based on publicly available fUS datasets, highlighting the framework's effectiveness in addressing data limitations.

\end{abstract}
\section{Introduction}

Functional ultrasound is a rapidly advancing neuroimaging modality distinguished by its exceptional spatiotemporal resolution, achieving micron-scale spatial precision and millisecond-level temporal accuracy~\cite{dizeux2019functional,renaudin2022functional,blaize2020functional,brunner2021whole}. By capitalizing on the principle of neurovascular coupling—the link between neural activity and cerebral hemodynamics—fUS facilitates high-fidelity, non-invasive investigation of brain function. Its applications are diverse, spanning from bedside neonatal monitoring~\cite{baranger2021bedside,demene2017functional,demene2016functional} and intraoperative surgical guidance~\cite{imbault2017intraoperative,soloukey2020functional,soloukey2024p09} to the development of next-generation brain-computer interfaces~\cite{griggs2024decoding,zheng2023emergence,agyeman2024functional}. However, the widespread clinical translation of fUS technology is impeded by two significant challenges: the chronic scarcity of human brain imaging data and the limitations of current generative models in synthesizing physiologically realistic fUS images.

The acquisition of large-scale human fUS datasets is fundamentally constrained by stringent ethical considerations and the substantial signal attenuation caused by the adult human skull. These factors not only limit the volume of available data but also curtail the demographic diversity within existing repositories. Concurrently, state-of-the-art generative methods, particularly diffusion models that excel at natural image synthesis, are ill-suited for fUS imaging. The iterative denoising process inherent to diffusion models progressively erodes the subtle hemodynamic signatures and fine-grained vascular architectures that are critical for an accurate representation of fUS data.

To address these challenges, a novel generative adversarial network~\cite{goodfellow2014generative} is proposed. It is tailored for high-fidelity fUS image synthesis. This framework features two primary architectural innovations: a Detail Feature Enhancement (DFE) module designed to preserve intricate microvascular textures and hemodynamic patterns, and a class-conditional batch normalization (cBatchNorm) layer that enforces physiological plausibility by conditioning the synthesis process on specific brain regions.

The main contributions of this paper are threefold:
\begin{enumerate}[label=\arabic*)]
\item We introduce a UNet-based discriminator architecture that performs hierarchical discrimination at multiple scales. This enables the generator to be optimized at a pixel level through concurrent analysis of both local and global features, significantly enhancing the reconstruction fidelity of microvascular networks and preserving subtle hemodynamic signals.
\item We strategically integrate a DFE module within the generator's encoder-decoder structure. This module employs multi-scale feature extraction to explicitly capture microvascular details at various resolutions, while its internal residual connections mitigate artifact generation and maintain the physiological integrity of the synthesized hemodynamic signals.
\item We propose the use of cBatchNorm to inject brain region-specific information directly into the normalization process. This allows our model to synthesize images that accurately reflect the distinct hemodynamic characteristics of targeted cortical areas, ensuring the anatomical and functional consistency of the generated data.
\end{enumerate}

\section{RELATED WORK}

\subsection{Functional Ultrasound Imaging}
Functional ultrasound is an emerging neuroimaging modality that enables the observation of cerebral dynamics and pathological processes with high spatiotemporal resolution in a non-invasive manner. 
Its clinical utility has been demonstrated in various contexts. 
For instance, Imbault et al.~\cite{imbault2017intraoperative} showcased the capability of fUS in detecting cortical responses to sensory stimuli during neurosurgery, highlighting its potential for real-time intraoperative monitoring. 
Pioneering work by Baranger et al.~\cite{baranger2021bedside} applied fUS to the bedside monitoring of functional connectivity in neonates, successfully identifying developmental distinctions between preterm and full-term infants at submillimeter resolution.
Despite its promise, the widespread adoption of fUS is hindered by several challenges, including the high cost of acquisition hardware and significant signal attenuation through the cranium. 
While efforts to democratize fUS data, such as the open-access dataset by Rabut et al.~\cite{rabut2024functional}, are crucial, existing datasets often suffer from limited sample sizes and a lack of demographic diversity.

\subsection{Medical Image Generation Algorithms}
Generative models, a cornerstone of modern artificial intelligence, have shown immense potential in revolutionizing medical imaging analysis and diagnostics~\cite{lei2021diagnosis,wang2025smart,wang2018bone,muller2023multimodal,zeng2017ga,mo2009variational,wang2024enhanced,wang2020ensemble,guo2025underwater,zuo2023alzheimer}.
Among these, Generative Adversarial Networks and Denoising Diffusion Probabilistic Models have become two of the most prominent paradigms, capable of synthesizing high-fidelity, disease specific medical images~\cite{zuo2024prior,pan2021characterization,dorjsembe2022three,zuo2021multimodal,zong2024new,li2023generative,jing2024estimating,zuo2023brain}.
A primary application of these models is to mitigate data scarcity by augmenting training datasets, thereby enhancing the performance of downstream tasks and improving image reconstruction quality.
Nevertheless, significant challenges persist, including ensuring the anatomical plausibility of generated images, accurately preserving critical pathological features, and achieving model robustness across diverse imaging protocols and patient populations.

\section{METHODOLOGY}

\subsection{Overview}
\begin{figure}[ht]
\includegraphics[width=\textwidth]{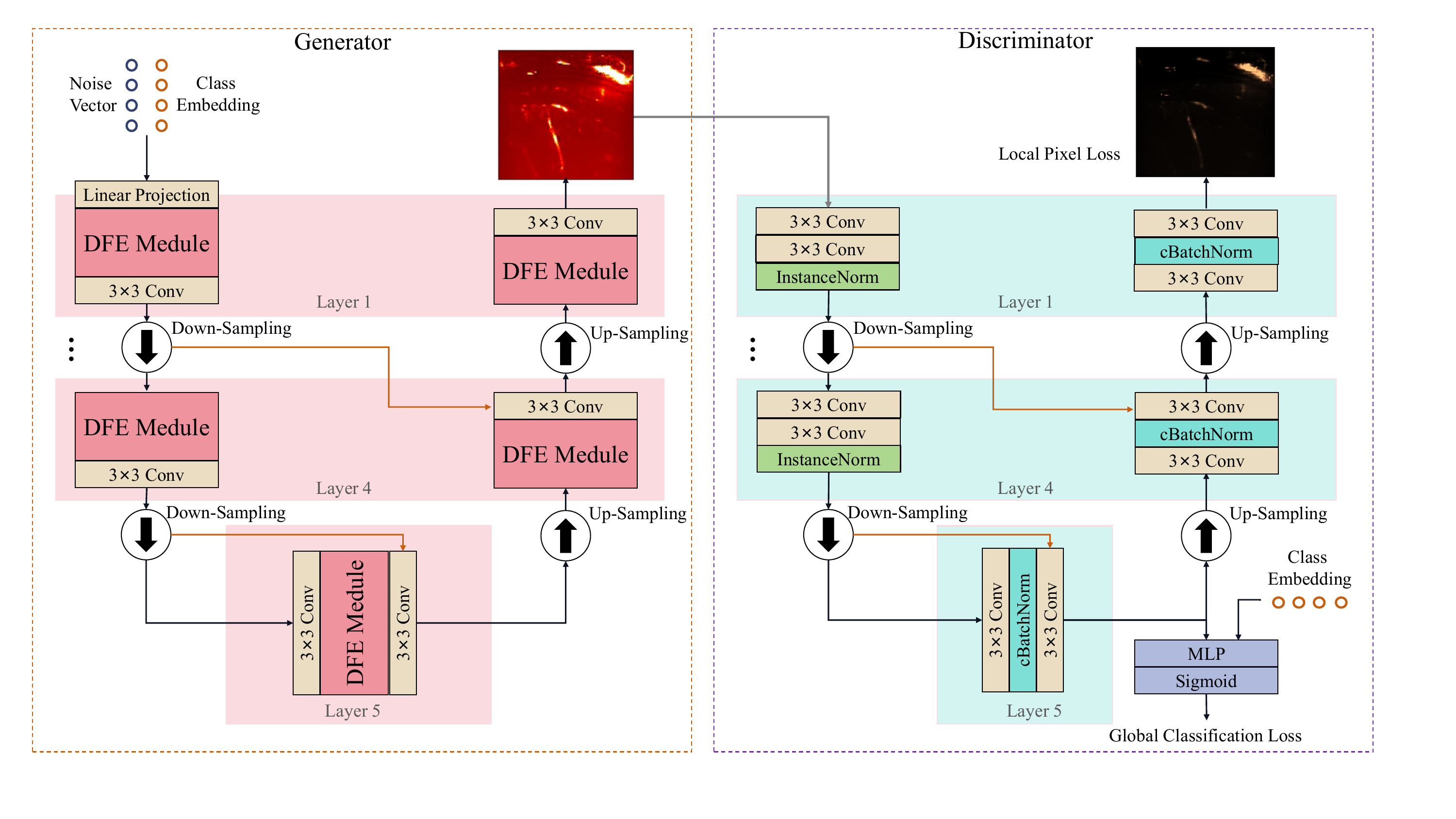}
\caption{The UltraDfeGAN framework employs a two-UNet architecture as its encoder-decoder backbone, enhanced through the incorporation of DEF modules and cBatchNorm modules.} 
\label{fig1}
\end{figure}

We introduce UltraDfeGAN, a multi-scale generative adversarial network designed to synthesize high-fidelity functional ultrasound images. As illustrated in Fig.~\ref{fig1}, our framework builds upon a U-Net-based adversarial architecture, enhanced with two primary contributions: a Detail Feature Enhancement module and a class-conditional Batch Normalization  layer. The DEF module employs multi-scale feature fusion to reconstruct fine-grained blood flow details with high precision. Concurrently, the cBatchNorm module integrates categorical metadata, such as task-specific brain states, by modulating the normalization parameters within the discriminator. This synergistic design enables UltraDfeGAN to generate images that not only capture intricate microvascular textures and dynamic hemodynamic signals but also maintain physiological plausibility conditioned on specific experimental paradigms.

\subsection{UltraDfeGAN Architecture}
The core of UltraDfeGAN is a generative adversarial network built upon a modified U-Net architecture for both the generator and the discriminator.

\textbf{Generator.} The generator network is designed to transform a random noise vector $z$, concatenated with a class embedding $c$, into a realistic fUS image. The input vector is first linearly projected and then processed by a cascade of DFE modules, which are organized in an encoder-decoder structure (see pink-shaded regions in Fig.~\ref{fig1}). The encoder path progressively downsamples the feature maps using $3\times3$ convolutions, capturing hierarchical features at multiple scales. After reaching the minimum resolution at the bottleneck layer, the decoder path employs transposed convolutions to upsample the feature maps. Skip connections between corresponding encoder and decoder layers facilitate the propagation of low-level spatial information, which is crucial for preserving fine-grained vascular details in the final generated image.

\textbf{Discriminator.} The discriminator (light-blue shaded regions in Fig.~\ref{fig1}) also adopts a U-Net structure to effectively distinguish between real and generated images at both global and local scales. Its encoder processes an input image to extract a hierarchy of features. These features are then utilized by two distinct heads: a global image classifier and a local patch discriminator.

The \textbf{global classifier} processes the feature map from the discriminator's bottleneck through a multi-layer perceptron (MLP) to produce a single scalar value representing the probability that the entire image is real. This encourages the generator to produce globally coherent and structurally plausible images. The corresponding adversarial loss is:
\begin{equation}
\mathcal{L}_{\text{global}} = -\mathbb{E}_x[\log D_{\text{global}}(x)] - \mathbb{E}_{z,c}[\log(1 - D_{\text{global}}(G(z,c)))].
\label{eq:global_loss}
\end{equation}

The \textbf{local patch discriminator} is formed by the decoder of the U-Net. It generates a probability map where each value corresponds to the realness of a patch in the input image. This forces the generator to synthesize realistic local textures and details. The loss is computed by aggregating the pixel-level predictions:
\begin{equation}
\mathcal{L}_{\text{local}} = -\mathbb{E}_x \left[ \sum_{i,j} \log [D_{\text{local}}(x)]_{i,j} \right] - \mathbb{E}_{z,c} \left[ \sum_{i,j} \log(1 - [D_{\text{local}}(G(z,c))]_{i,j}) \right].
\label{eq:local_loss}
\end{equation}

The total discriminator loss is the sum of the global and local losses: $\mathcal{L}_D = \mathcal{L}_{\text{global}} + \mathcal{L}_{\text{local}}$. Correspondingly, the generator is trained to minimize an analogous loss, $\mathcal{L}_G$, compelling it to simultaneously optimize for both global structural integrity and local photorealism.

\subsection{Detail Feature Enhancement Module}
\begin{figure}[ht]
\includegraphics[width=\textwidth]{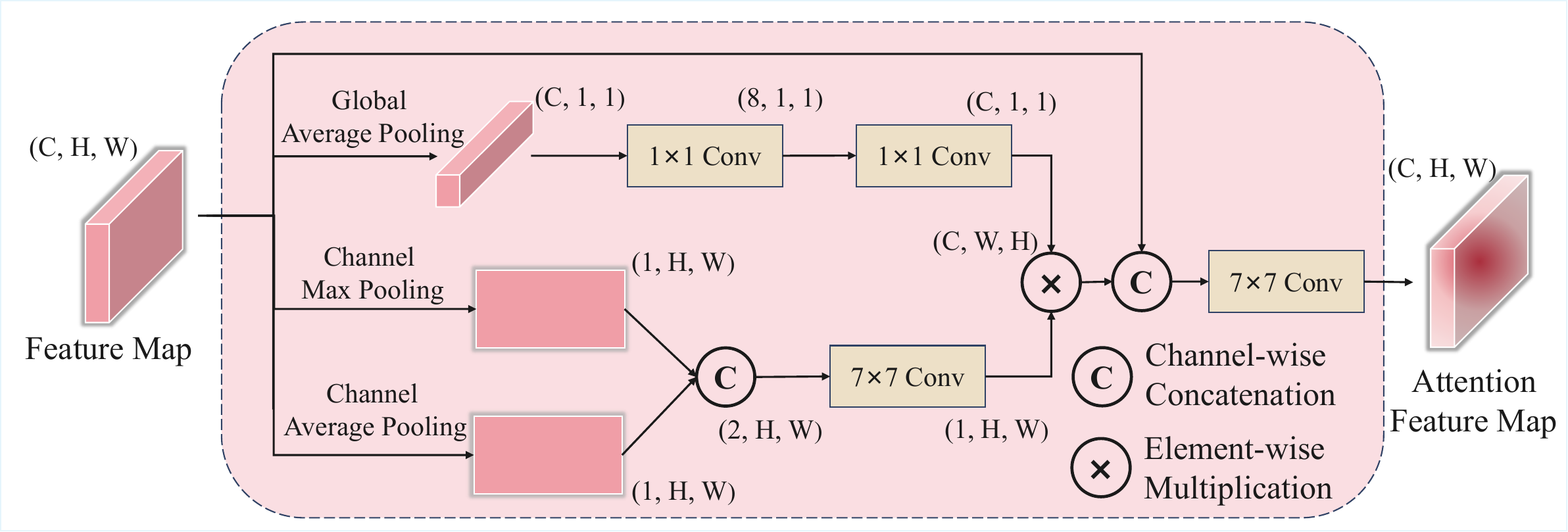}
\caption{The DEF module architecture diagram presents the integration of multi-scale feature fusion and dynamic attention mechanisms, designed to enhance structural detail preservation and physiological coherence in functional ultrasound imaging.} 
\label{fig2}
\end{figure}

The DEF module is engineered to preserve fine-grained vascular details and suppress noise artifacts through a multi-scale feature fusion mechanism (see Fig.~\ref{fig2}). The module takes an input feature map and processes it through three parallel branches to capture multi-level feature statistics.

The first branch applies global average pooling (GAP) to compress the spatial dimensions, producing a channel-wise descriptor that encapsulates global context, such as overall hemodynamic patterns. The other two branches employ channel-wise max pooling and average pooling, respectively, to extract salient local features and mean intensity information.

The outputs of these branches are transformed via shared $1\times1$ convolutions and nonlinear activation functions to model complex feature dependencies. These parallel representations are then concatenated and processed to generate a dynamic channel attention map. This map is subsequently applied to the original input feature map via element-wise multiplication, adaptively recalibrating channel-wise features to emphasize important vascular signals while attenuating irrelevant noise. Finally, a residual connection is employed to sum the attention-modulated features with the original input, ensuring stable training and preserving learned topological relationships in the output feature map.

\subsection{Class-Conditional Batch Normalization}
To efficiently condition the generation process on categorical metadata without the computational overhead and potential mode collapse associated with direct feature concatenation, we incorporate a cBatchNorm layer within the discriminator. This approach allows the network to modulate its behavior based on experimental conditions, such as different stimuli or brain states.

Within each cBatchNorm layer, the feature map $x$ is first normalized using its batch-derived mean $\mu$ and standard deviation $\sigma$. The normalized activation is then modulated by a class-specific affine transformation:
\begin{equation}
\hat{x} = \gamma(c) \cdot \frac{x - \mu}{\sqrt{\sigma^2 + \epsilon}} + \beta(c),
\end{equation}
where $\epsilon$ is a small constant for numerical stability. The scaling parameter $\gamma(c)$ and shifting parameter $\beta(c)$ are dynamically generated by feeding the class embedding $c$ into two separate, lightweight MLPs. This mechanism enables the discriminator to learn class-specific feature representations, thereby guiding the generator to produce images that are faithful to the given condition.

\section{EXPERIMENTS}

\subsection{Dataset and Preprocessing}
We conduct our experiments on the public fUS dataset introduced by Rabut et al.~\cite{rabut2024functional}. This dataset comprises cerebral hemodynamic responses acquired from the human brain under two distinct conditions: a motor task (piano playing), a cognitive task (Line-connecting), and a baseline resting state. Each experimental session is structured with 50-second task blocks alternating with 100-second resting intervals. The fUS acquisitions targeted key cortical areas, including the primary motor cortex, primary somatosensory cortex, and posterior parietal cortex.

Following the original study's protocol, the dataset is partitioned into training and test sets. The piano-playing task subset contains 720 training and 270 test images, while the connect-the-dots task subset consists of 1,530 training and 270 test images. To ensure computational efficiency and standardize the model input, all fUS images were uniformly cropped to a resolution of $128\times128$ pixels.

\subsection{Baseline Models and Implementation Details}
\begin{figure}[ht]
\includegraphics[width=\textwidth]{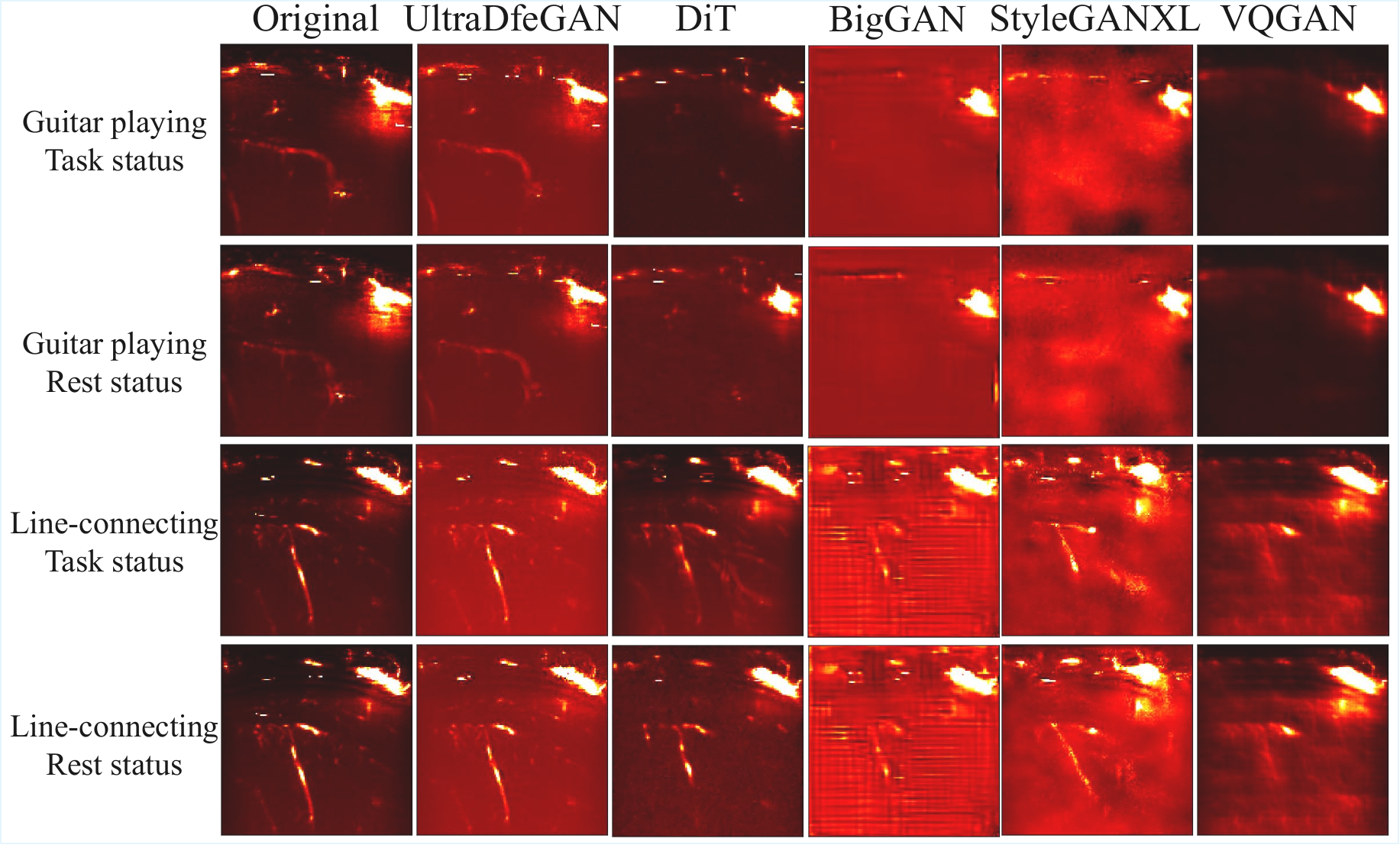}
\caption{This figure presents a comparative analysis of generative performance across different models under guitar-playing and Line-connecting tasks in both task and resting-state scenarios. The first column displays real images, followed by generated results from UltraDfeGAN, DiT, BigGAN, StyleGAN-XL, and VQGAN.} 
\label{fig3}
\end{figure}

\subsection{Experimental Setup}
\label{sec:implementation_details}
We benchmark our proposed model, UltraDfeGAN, against several prominent generative models: DiT~\cite{peebles2023scalable}, BigGAN~\cite{brock2018large}, StyleGAN-XL~\cite{sauer2022stylegan}, and VQGAN~\cite{yu2021vector}. All models were implemented using the PyTorch framework and trained on NVIDIA A800 80GB PCIe GPUs. 

For the training of UltraDfeGAN, we employed the Adam optimizer for both the generator and the discriminator. The learning rate was set to $2 \times 10^{-4}$, with momentum parameters $\beta_1 = 0.5$ and $\beta_2 = 0.999$. It is noteworthy that no data augmentation techniques were applied during the training process, ensuring a fair comparison of the models' intrinsic generative capabilities.

\subsection{Comparative Analysis of Image Generation}
A comprehensive comparison against baseline models substantiates the superior performance of UltraDfeGAN in synthesizing high-fidelity fUS images. As illustrated qualitatively in Fig.~\ref{fig3}, our model generates vascular patterns for the piano-playing task that are visually akin to the ground truth. The resulting images exhibit continuous microvascular structures and sharply defined flow signals. In contrast, competing methods such as DiT, BigGAN, and VQGAN produce images with noticeable vascular discontinuities and noise artifacts. Furthermore, under resting-state conditions, UltraDfeGAN excels at suppressing background noise, a challenge for models like StyleGAN-XL, which introduces discernible checkerboard artifacts.

\begin{figure}[ht]
\includegraphics[width=\textwidth]{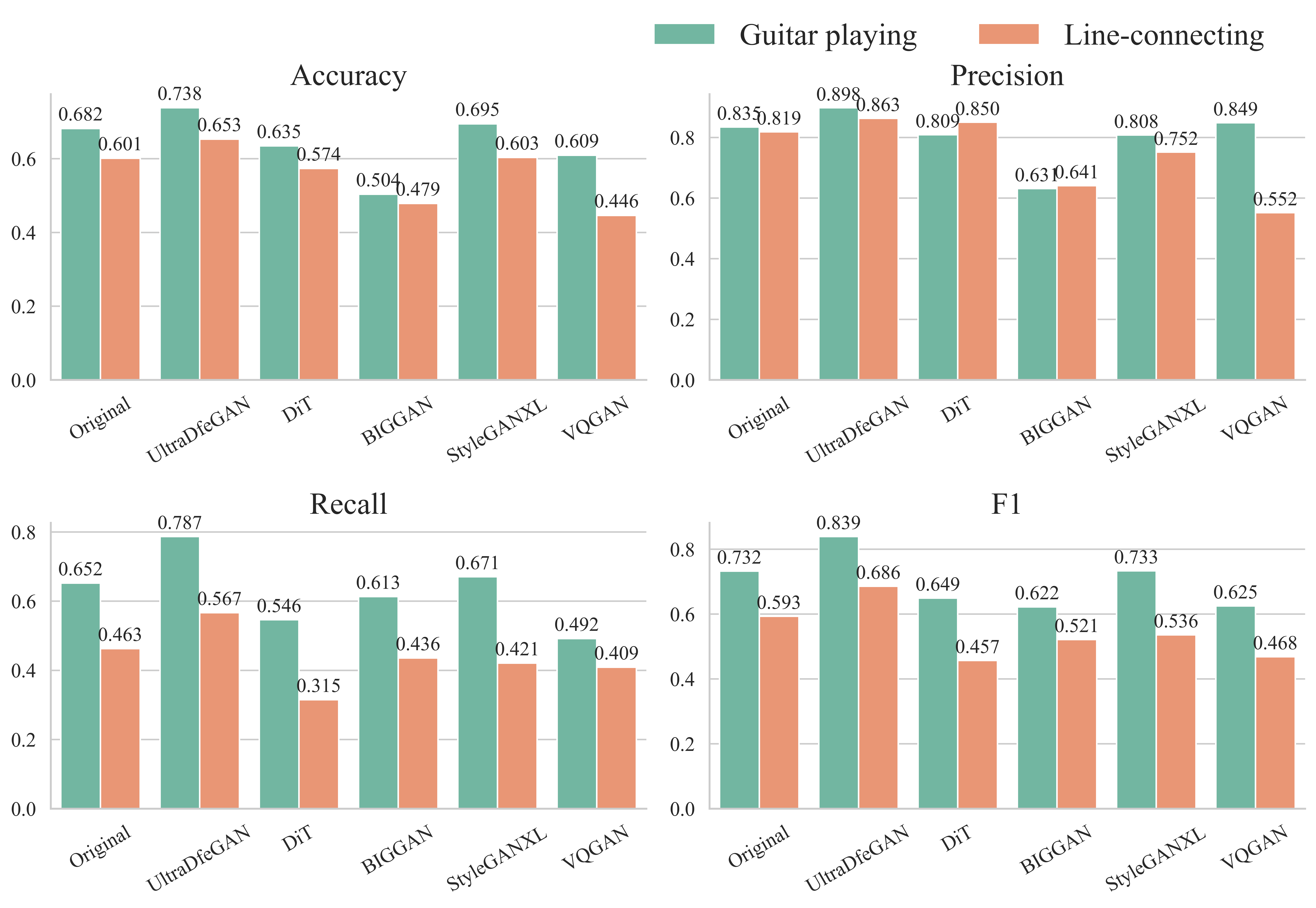}
\caption{A quantitative comparison of image quality metrics across generative models further demonstrates that UltraDfeGAN consistently outperforms all baseline models in terms of quantitative evaluation metrics.} 
\label{fig4}
\end{figure}

The quantitative results, presented in Fig.~\ref{fig4}, corroborate these qualitative observations. For the piano-playing task, UltraDfeGAN achieves an SSIM of 0.85 and a multi-scale SSIM (MS-SSIM) of 0.96, markedly surpassing DiT. Notably, its Fréchet Inception Distance (FID) of 92.95 represents a 51.5\% improvement over BigGAN. Our model consistently outperforms all baselines across all reported metrics on the connect-the-dots task, further affirming its state-of-the-art performance.

\subsection{Evaluation on Downstream Task Performance}
To assess the practical utility of the synthesized data, we evaluated its impact on a downstream task: task-state classification from fUS images. We employed a Principal Component Analysis (PCA) followed by a Random Forest classifier. This classifier was trained on two distinct datasets: (1) the original training set and (2) a version augmented with images generated by UltraDfeGAN. 

\begin{figure}[ht]
\includegraphics[width=\textwidth]{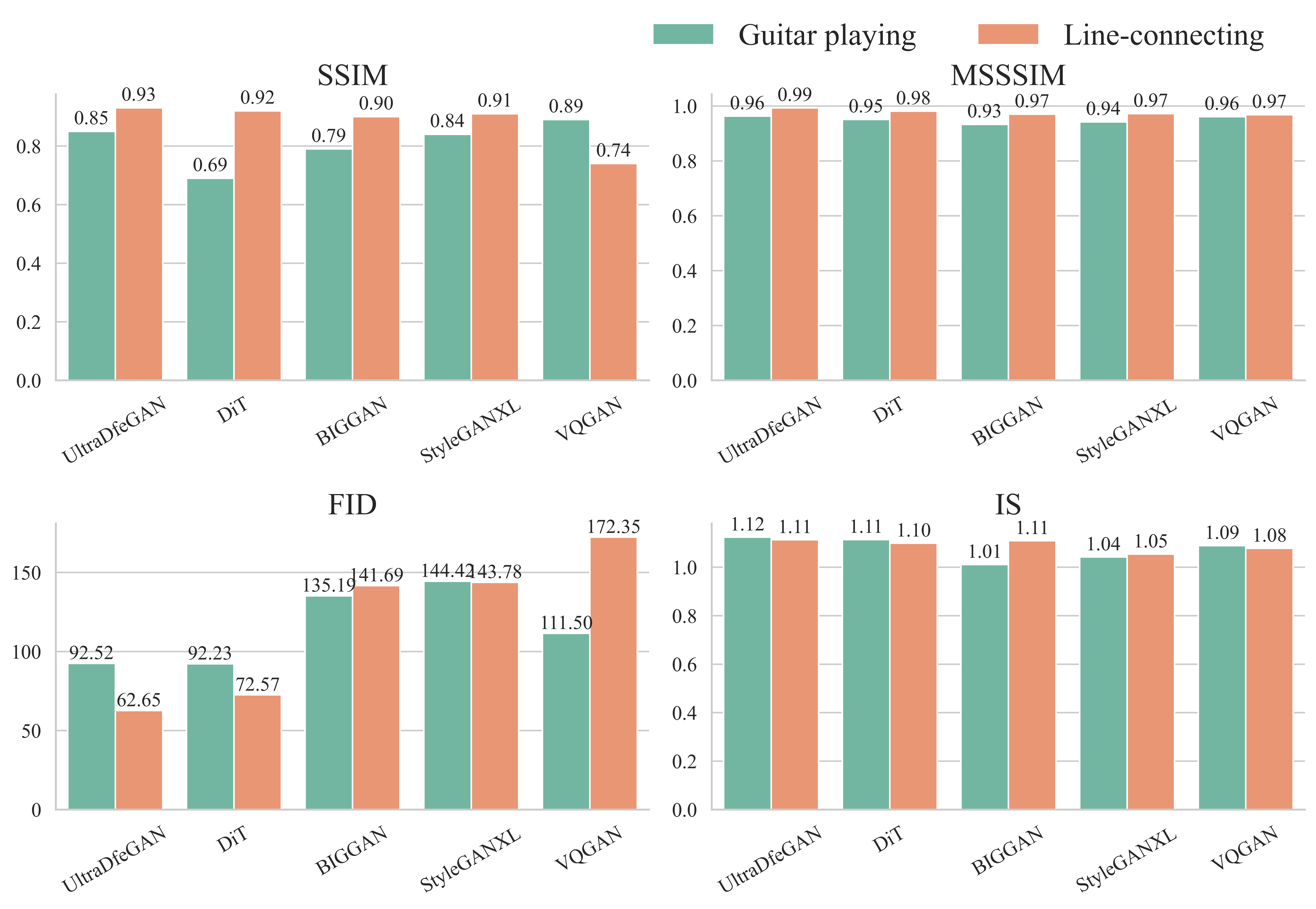}
\caption{Quantitative Evaluation of Downstream Classification Task Performance.} 
\label{fig5}
\end{figure}

As depicted in Fig.~\ref{fig5}, augmenting the training data with our synthesized samples yields a significant improvement in classification accuracy. This result demonstrates that UltraDfeGAN not only generates visually realistic images but also captures salient features of the underlying neural activity, which are beneficial for downstream analytical tasks.

\subsection{Ablation Studies}
To validate the contributions of the core components of UltraDfeGAN, we conducted a series of ablation studies. In these experiments, we systematically reverted our proposed architectural modifications to the original UNetGAN~\cite{schonfeld2020u} structure and evaluated the impact on performance. The evaluation was based on the quantitative metrics from the downstream classification task.

The results, summarized in Table~\ref{tab:performance_metrics}, unequivocally demonstrate that each proposed module contributes significantly to the model's overall efficacy. The removal of either component resulted in a marked degradation across key performance metrics. This confirms the critical role of our architectural designs in achieving the reported state-of-the-art results.
\begin{table}[ht]
    \centering
    \caption{Ablation studies for guitar playing and line-connecting tasks.}
    \label{tab:performance_metrics}
    \begin{tabular}{c c c c c c c c c}
        \toprule
        & \multicolumn{4}{c}{Guitar playing} & \multicolumn{4}{c}{Line-connecting} \\
        \cmidrule(r){2-5} \cmidrule(l){6-9}
        & Accuracy & Precision & Recall & F1-Score & Accuracy & Precision & Recall & F1-Score \\
        \midrule
        Original & 0.682 & 0.835 & 0.652 & 0.732 & 0.601 & 0.819 & 0.436 & 0.593 \\
        w/o DEF & 0.695 & 0.845 & 0.67 & 0.752 & 0.615 & 0.83 & 0.45 & 0.61 \\
        w/o cBatch & 0.72 & 0.86 & 0.685 & 0.77 & 0.63 & 0.845 & 0.47 & 0.635 \\
        Proposed & 0.738 & 0.898 & 0.787 & 0.839 & 0.653 & 0.863 & 0.567 & 0.686 \\
        \bottomrule
    \end{tabular}
\end{table}

\section{Conclusion}
This study addresses the challenges of data scarcity and model fairness in the clinical translation of functional ultrasound imaging by proposing a novel GAN-based data augmentation framework, UltraDfeGAN. The core contributions include: (1) Introducing a U-Net discriminator with a pixel-level-global joint optimization mechanism into ultrasound image generation, achieving submillimeter-resolution reconstruction of microvascular textures; (2) Developing the Detail Feature Enhancement module, which employs multi-scale feature fusion and dynamic attention mechanisms to suppress noise while preserving physiological plausibility of hemodynamic signals; (3) Incorporating cBatchNorm modules that embed prior knowledge of brain region activity states into the generation process, significantly enhancing cross-task generalization capability. Experimental results demonstrate that this framework not only outperforms existing models on quantitative metrics but also exhibits substantial improvements in practical utility for downstream tasks. Future work will focus on cross-modal data fusion and robustness optimization of generative models to advance the application of functional ultrasound imaging in neuromodulation.

% \begin{credits}
% \subsubsection{\ackname} A bold run-in heading in small font size at the end of the paper is
% used for general acknowledgments, for example: This study was funded

\subsubsection{\discintname}
% It is now necessary to declare any competing interests or to specifically
% state that the authors have no competing interests. Please place the
% statement with a bold run-in heading in small font size beneath the
% (optional) acknowledgments\footnote{If EquinOCS, our proceedings submission
% system, is used, then the disclaimer can be provided directly in the system.},
The authors have no competing interests to declare that are relevant to the content of this article. 
%Or: Author A has received research
% grants from Company W. Author B has received a speaker honorarium from
% Company X and owns stock in Company Y. Author C is a member of committee Z.
% \end{credits}
\bibliographystyle{splncs04} % 或 plain
\bibliography{ref.bib}
\end{document}